\documentclass[12pt]{article}

\usepackage{latexsym}
\usepackage{amsmath,amsfonts}
\usepackage{times}
\allowdisplaybreaks[4]

\catcode`\@=11

\newcount\hour
\newcount\minute
\newtoks\amorpm \hour=\time\divide\hour by 60\minute
=\time{\multiply\hour by 60 \global\advance\minute by-\hour}
\edef\standardtime{{\ifnum\hour<12 \global\amorpm={am}%
        \else\global\amorpm={pm}\advance\hour by-12 \fi
        \ifnum\hour=0 \hour=12 \fi
        \number\hour:\ifnum\minute<10
        0\fi\number\minute\the\amorpm}}
\edef\militarytime{\number\hour:\ifnum\minute<10 0\fi\number\minute}

\def\draftlabel#1{{\@bsphack\if@filesw {\let\thepage\relax
   \xdef\@gtempa{\write\@auxout{\string
      \newlabel{#1}{{\@currentlabel}{\thepage}}}}}\@gtempa
   \if@nobreak \ifvmode\nobreak\fi\fi\fi\@esphack}
        \gdef\@eqnlabel{#1}}
\def\@eqnlabel{}
\def\@vacuum{}
\def\marginnote#1{}
\def\draftmarginnote#1{\marginpar{\raggedright\scriptsize\tt#1}}
\def\draft{
        \pagestyle{plain}
        \overfullrule=2pt
        \oddsidemargin -.5truein
        \def\@oddhead{\sl \phantom{\today\quad\militarytime} \hfil
        \smash{\Large\sl DRAFT} \hfil \today\quad\militarytime}
        \let\@evenhead\@oddhead
        \let\label=\draftlabel
        \let\marginnote=\draftmarginnote
        \def\ps@empty{\let\@mkboth\@gobbletwo
        \def\@oddfoot{\hfil \smash{\Large\sl DRAFT} \hfil}
        \let\@evenfoot\@oddhead}
        \def\@eqnnum{(\theequation)\rlap{\kern\marginparsep\tt\@eqnlabel}%
        \global\let\@eqnlabel\@vacuum}  }

\newcommand{\rf}[1]{(\ref{#1})}
\renewcommand{\theequation}{\thesection.\arabic{equation}}
\renewcommand{\thefootnote}{\fnsymbol{footnote}}
\newcommand{\newsection}{   
\setcounter{equation}{0}\section}

\def\appendix#1{\addtocounter{section}{1}\setcounter{equation}{0}
\renewcommand{\thesection}{\Alph{section}}
\section*{Appendix \thesection\protect\indent \parbox[t]{11.15cm}{#1}}
\addcontentsline{toc}{section}{Appendix \thesection\ \ \ #1}}

\def\be{\begin{equation}}
\def\ee{\end{equation}}
\def\beq{\begin{eqnarray}}
\def\eeq{\end{eqnarray}}

\def\parline{\,\partial\kern -0.55em /\,\,}

\def\half{{\frac{1}{2}}}

\def\DD{{\cal D}}

\def\LL{{\cal L}}
\def\MM{{\cal M}}
\def\NN{{\cal N}}

\def\TT{{\cal T}}

\def\Dbf{{\bf D}}

\def\Gbf{{\bf G}}

\def\Lbf{{\bf L}}

\def\ebf{{\bf e}}
\def\mbf{{\bf m}}

\def\alphabf{{\boldsymbol{\alpha}}}

\def\mubf{{\boldsymbol{\mu}}}
\def\Pibf{{\boldsymbol{\Pi}}}

\def\alphabf{{\boldsymbol{\alpha}}}
\def\mubf{{\boldsymbol{\mu}}}
\def\Pibf{{\boldsymbol{\Pi}}}

\def\phik{|\phi\rangle}
\def\phibr{\langle\phi|}
\def\ck{|c\rangle}
\def\cbk{|\bar{c}\rangle}
\def\cbr{\langle c |}
\def\cbbr{\langle \bar{c}|}
\def\bk{|b\rangle}
\def\bbr{\langle b |}

\def\xik{|\xi\rangle}

\def\smponetwo{{\scriptscriptstyle [1,2]}}

\def\alpar{\alpha\partial}
\def\albpar{{\bar\alpha\partial}}

\def\kappab{{\bar{\kappa}}}

\def\Lb{\bar{L}}

\def\cb{\bar{c}}
\def\eb{\bar{e}}

\def\(A)dS{{\rm (A)dS}}
\def\pAdS{{\rm (A)dS}}

\def\eff{{\rm eff}}
\def\tot{{\rm tot}}

\def\sh{{\rm sh}}

\def\FP{{\rm FP}}

\def\qu{{\rm qu}}
\def\I{{\rm I}}
\def\II{{\rm II}}
\def\lin{{\rm lin}}

\def\ssf{{\sf s}}
\def\ssfb{\bar{\sf s}}

\def\mun{{\underline{m}}}

\newcommand{\mc}{\multicolumn}

\begin{document}


\begin{flushright}
FIAN-TD-2014-11 \qquad \ \ \ \ \ \ \  \\
arXiv: 1407.2601 [hep-th] \ \ \ \
\end{flushright}

\vspace{1cm}

\begin{center}

{\Large \bf BRST invariant effective action of shadow fields, conformal

 \bigskip

 fields, and AdS/CFT }

\vspace{2.5cm}

R.R. Metsaev%
\footnote{ E-mail: metsaev@lpi.ru
}

\vspace{1cm}

{\it Department of Theoretical Physics, P.N. Lebedev Physical
Institute, \\ Leninsky prospect 53,  Moscow 119991, Russia }

\vspace{3.5cm}

{\bf Abstract}

\end{center}

\bigskip

Totally symmetric arbitrary spin massless and massive fields in AdS space are studied. For such fields, we obtain Lagrangians which are invariant under global BRST transformations. The Lagrangians are used for computation of partition functions and effective actions. We demonstrate that BRST invariant bulk action for massless field evaluated on the solution of Dirichlet problem for gauge massless fields and Faddeev-Popov fields leads to BRST invariant effective action for canonical shadow gauge fields and shadow Faddeev-Popov fields, while the BRST invariant bulk action for massive field evaluated on the solution of Dirichlet problem for gauge massive fields and Faddeev-Popov fields leads to BRST invariant effective action for anomalous shadow gauge fields and shadow Faddeev-Popov fields. The leading logarithmic divergence of the regularized effective action for the canonical shadow field leads to simple BRST invariant action of conformal field. We demonstrate that the Nakanishi-Laudrup fields entering the BRST invariant Lagrangian of conformal field can geometrically be interpreted as boundary values of massless AdS fields.

\newpage
\renewcommand{\thefootnote}{\arabic{footnote}}
\setcounter{footnote}{0}

\newsection{ Introduction }

Slavnon-Taylor identities \cite{Slavnov:1972fg} and BRST approach \cite{Becchi:1975nq} play important role in studies of the renormalizations of gauge theories. In the quantum field theory, Heisenberg equations of motion are not well defined. Therefore, in studies of various theories of  interacting quantized gauge fields, usually one concentrates on the study of S-matrix.  In such studies, it is the Slavnov-Taylor identities and BRST approach that allow to carry out the renormalization procedure in a relativistic invariant fashion to obtain unitary renormalized S-matrix. Often, the unitary S-matrix is not obtained by using a priori prescribed computation procedure.  This is to say that the unitary S-matrix is often obtained by using the self-consistent computational framework. In some sense one can say that it is the procedure for the derivation of the unitary S-matrix that provides us the constructive way for building theory of quantized interacting fields.

At the present time, a notion of the S-matrix is more or less well understood for theories of quantized fields in flat space. Generalization of notion of the  S-matric to quantized fields in AdS space seems to be problematic.  On the other hand, as was pointed out somewhere in earlier literature, the role of the S-matric for fields in AdS space can be delegated to so called effective action of shadow fields. Such action is defined through a continual integral over AdS field with some particular boundary condition to be imposed on boundary value of AdS field. That particular boundary value of AdS field is referred to as shadow field. Therefore the continual integral provides the realization of the effective action in terms of the shadow field. In the framework of AdS/CFT correspondence, the effective action is realized as a generating function for correlation functions of CFT which, according to Maldacena conjecture, is dual to AdS theory.  By analogy with the S-matrix in flat space, one can believe that the full quantum effective action of shadow fields provides the constructive way for a definition of quantum field theory in AdS space. For these reasons, study of the effective action of shadow fields seems to be well motivated.

In quadratic approximation, the effective actions for spin-1 and spin-2 shadow fields were studied in Refs.\cite{Freedman:1998tz}-\cite{Metsaev:2010zu},  while the effective action for arbitrary spin-$s$ shadow field was considered in Refs.\cite{Metsaev:2009ym,Metsaev:2011uy}. Some group-theoretical issues related to problems of the computation of the effective actions were addressed in Refs.\cite{Dobrev:1998md}-\cite{Metsaev:2008fs}. Effective action for arbitrary spin shadow field found in Refs.\cite{Metsaev:2009ym,Metsaev:2011uy} was presented in terms of gauge field subject to some differential constraints.  In Sec. 3, we introduce shadow Faddeev-Popov fields and find BRST invariant effective action expressed in terms of fields which are not subject to any differential constraints.

Effective actions of shadow fields that are dual to massless AdS fields has logarithmic divergences. For the case of spin-2 field in Ref.\cite{Liu:1998bu} and arbitrary spin-$s$ field in Ref.\cite{Metsaev:2009ym}, it was demonstrated that the logarithmic divergences of the effective actions of spin-2 and arbitrary spin-$s$ shadow fields turn out to be actions of the respective conformal spin-2 and arbitrary spin-$s$ conformal fields. In Sec. 4, we show that logarithmic divergence of the BRST invariant action of arbitrary spin-$s$ canonical shadow field turns out to be BRST invariant action of arbitrary spin-$s$ conformal field. In due course, we demonstrate that Nakanishi-Laudrup fields entering BRST invariant action of conformal field can geometrically be interpreted as boundary values of massless AdS fields.

In recent time, some attention has been paid to the problem of the computation of one-loop partition function for AdS fields (see, e.g., Refs.\cite{Gupta:2012he,Tseytlin:2013jya} and references therein). As BRST invariant  Lagrangian for AdS fields provides systematical and self-contained way for the computation of partition function we start our discussion in Sec. 2 with building  BRST invariant Lagrangian for AdS fields and apply such Lagrangian for the computation of partition function.

\newsection{  $so(d,1)$ covariant approach and partition function }

In this section, we review the Lagrangian $so(d,1)$ covariant formulation of dynamics of free fields in  (A)dS${}_{d+1}$ space developed in Refs.\cite{Metsaev:2008ks,Metsaev:2009hp}. In Sec.\ref{secpartfun}, we will use our approach for the derivation of BRST invariant Lagrangians of (A)dS fields and apply such the Lagrangians for the study of partition functions.

\noindent {\bf Field content}. For the gauge invariant description of spin-$s$ massless field in  (A)dS$_{d+1}$, we use a double-traceless tensor field of the $so(d,1)$ algebra,
\be \label{21062014-man-01}
\phi^{A_1\ldots A_s}\,, \qquad \phi^{AABBA_5\ldots A_s}=0,  \quad \hbox{ for } s \geq 4\,,
\ee
while, for the gauge invariant description of spin-$s$ massive field in  AdS$_{d+1}$, we use the following scalar, vector, and totally symmetric double-traceless tensor fields of the  $so(d,1)$ algebra:
\be \label{21062014-man-02}
\phi^{A_1\ldots A_{s'}} \,, \quad s'=0,1,\ldots, s\,, \qquad \phi^{AABBA_5\ldots A_{s'}}=0 \quad \hbox{ for } s' \geq 4\,.
\ee
We note that field contents in \rf{21062014-man-01}, \rf{21062014-man-02} enter so called metric-like formulation of massless and massive fields.%
\footnote{ In the framework of metric-like approach, massless fields in AdS${}_4$ were studied in Ref.\cite{Fronsdal:1978vb}, while massless fields in AdS${}_{d+1}$, $d\geq 3$, were considered in Refs.\cite{Metsaev:1999ui,Buchbinder:2001bs}. For massive field, field content in \rf{21062014-man-02}, was introduced in  \cite{Zinoviev:2001dt}. In the framework of frame-like approach, massless and massive fields were studied in Refs.\cite{Lopatin:1987hz}.}
Also we note that the $so(d,1)$ covariant formulation is developed by using an arbitrary parametrization of (A)dS space.

The presentation of gauge-invariant Lagrangian can considerably be simplified by using generating form of gauge fields. To this end we introduce oscillators $\alpha^A$, $\zeta$. Using such oscillators, gauge fields in \rf{21062014-man-01}, \rf{21062014-man-02} can be collected into a ket-vector $\phik$ as follows
\beq
\label{21062014-man-03} && \phik = \frac{1}{s!}  \alpha^{A_1} \ldots \alpha^{A_s}  \phi^{A_1\ldots A_s}|0\rangle,\hspace{2cm} \hbox{ for massless field};
\\
\label{21062014-man-04} && \phik = \sum_{s'=0}^{s} \frac{\zeta^{s-s'}}{\sqrt{(s-s')!}} |\phi^{s'}\rangle \,,
\nonumber\\
&& |\phi^{s'}\rangle = \frac{1}{s'!}  \alpha^{A_1} \ldots \alpha^{A_{s'}}  \phi^{A_1\ldots A_{s'}}|0\rangle\,, \hspace{1.5cm}  \hbox{ for massive field}.
\eeq

\noindent{\bf Gauge invariant Lagrangian}.
In terms of the ket-vector $\phik$, gauge invariant Lagrangians for massless and massive fields can be presented on an equal footing as follows,
\beq
\label{21062014-man-05} \LL  &  = &  \half  e \phibr  E \phik \,,
\\
\label{21062014-man-06} E & = & \mubf \bigl(\Box_{_\pAdS} + \mbf_1 + \rho
\alphabf^2\bar\alphabf^2\bigr) - \Lbf \bar\Lbf\,,
\\
\label{21062014-man-06x1} && \bar\Lbf \equiv  \bar\alphabf \Dbf - \half \alphabf \Dbf  \bar\alphabf^2  -
\bar\ebf_1\Pibf^\smponetwo + \half \ebf_1 \bar\alphabf^2\,,
\\
\label{21062014-man-06x2} && \Lbf \equiv \alphabf \Dbf  - \half \alphabf^2 \bar\alphabf \Dbf  - \ebf_1 \Pibf^\smponetwo + \half \bar\ebf_1 \alphabf^2\,,
\eeq
where $\rho =\epsilon/R^2$, $\epsilon=1(-1)$ for dS (AdS), $R$ is radius of (A)dS,  $e=\det e_\mun^A$, $e_\mun^A$ are vielbein in (A)dS, $\Box_{\pAdS}$  D'Alembert operator in (A)dS, while $\mbf_1$, $\ebf_1$, $\bar\ebf_1$ are given by
\beq
\label{21062014-man-06x3} &&  \hspace{-0.9cm}  \mbf_1 = \rho \bigl( s(s+d-5) -2d+ 4\bigr), \qquad
\ebf_1 =   0, \qquad \bar\ebf_1 = 0, \hspace{0.5cm} \hbox{ for massless field},
\\
\label{21062014-man-06x4} &&   \hspace{-0.9cm} \mbf_1 = -m^2 +\rho \Bigl( s(s+d-5) -2d+ 4 +N_\zeta(2s+d-1-N_\zeta)\Bigr)\,,
\nonumber\\
&&  \hspace{-0.9cm}  \ebf_1 =   \zeta \ebf_\zeta \,, \qquad \bar\ebf_1 = -   \ebf_\zeta \bar\zeta\,, \qquad
\nonumber\\
&&  \hspace{-0.9cm} \ebf_\zeta \equiv \Bigl(\frac{2s+d-3-N_\zeta}{2s+d-3-2N_\zeta} \bigl( m^2
-\rho N_\zeta (2s+d-4-N_\zeta)\bigr)\Bigr)^{1/2},   \hbox{ for massive field}.\qquad
\eeq
Operators $\alphabf \Dbf$, $\alphabf^2$, $\mubf$, $\Pibf^\smponetwo$ , $N_\zeta$ appearing in \rf{21062014-man-06}-\rf{21062014-man-06x4} are given in Appendix. Our bra-vectors are defined as $\phibr \equiv (\phik)^\dagger$. We emphasize that, in our approach, it is the use of the operators $\Lbf$, $\bar\Lbf$ \rf{21062014-man-06x1}, \rf{21062014-man-06x2} that simplifies considerably the structure of Lagrangian.%
\footnote{ Representation for Lagrangian in  \rf{21062014-man-05}-\rf{21062014-man-06x2} was found in Refs.\cite{Metsaev:2008ks,Metsaev:2009hp}. Other representations for the Lagrangian may be found in Refs.\cite{Buchbinder:2001bs,Zinoviev:2001dt}.}
Also we note the  helpful relation $e \phibr \Lbf\bar\Lbf \phik =  - e \langle \bar\Lbf \phi| \bar\Lbf \phik$ which is valid up to total derivatives. Note that $ \langle \bar\Lbf \phi| \equiv (\bar\Lbf \phik)^\dagger$.

Taking into account that the tensor fields entering our ket-vector $\phik$ are double-traceless, $(\bar\alphabf^2)^2\phik=0$, we note that, alternatively, operator $E$ \rf{21062014-man-06} can be represented as
\be
E  =  \Box_\pAdS + M_1 -  \frac{1}{4}\alphabf^2 \bar\alphabf^2 ( \Box_\pAdS + M_2)  - \Lbf \bar\Lbf\,,
\ee
where  $M_1$, $M_2$ are given by
\beq
&& \hspace{-1.9cm} M_1 = \rho \bigl( s(s+d-5) -2d+ 4) \bigr)\,,
\nonumber\\
&& \hspace{-1.9cm} M_2 = \rho \bigl( s(s+d-1) - 6\bigr)\,, \hspace{5.6cm} \hbox{ for massless field;}
\\
&& \hspace{-1.9cm} M_1 = -m^2 +\rho \Bigl( s(s+d-5) -2d+ 4 +N_\zeta(2s+d-1-N_\zeta)\Bigr)\,,
\nonumber\\
&& \hspace{-1.9cm} M_2 = -m^2 +\rho \Bigl( s(s+d-1) - 6 + N_\zeta(2s+d-5-N_\zeta)\Bigr)\,, \hspace{0.3cm} \hbox{ for massive field.}
\eeq

\noindent {\bf Gauge symmetries}. For the description of gauge symmetries of the massless spin-$s$ field, we use totally symmetric traceless rank-$(s-1)$ tensor field of the $so(d,1)$ algebra,
\be \label{25062014-man-01}
\xi^{A_1\ldots A_{s-1}}\,, \qquad \xi^{AAA_3\ldots A_{s-1}}=0\,, \quad \hbox{ for } s \geq 3,
\ee
while, for the description of gauge symmetries of the massive spin-$s$ field, we use the following scalar, vector, and totally symmetric traceless tensor fields of the $so(d,1)$ algebra:
\be \label{25062014-man-02}
\xi^{A_1\ldots A_{s'}} \,, \quad s'=0,1,\ldots, s-1\,, \qquad \xi^{AAA_3\ldots A_{s'}}=0 \quad \hbox{ for } s' \geq 2\,.
\ee

The presentation of gauge transformations can considerably be simplified by using the generating form of the scalar, vector, and tensor fields in \rf{25062014-man-01}, \rf{25062014-man-02},
\beq
\label{25062014-man-03} && \xik = \frac{1}{(s-1)!}  \alpha^{A_1} \ldots \alpha^{A_{s-1}}  \xi^{A_1\ldots A_{s-1}}|0\rangle\,, \hspace{1cm} \hbox{ for massless field};
\\
\label{25062014-man-04} && \xik = \sum_{s'=0}^{s-1} \frac{\zeta^{s-1-s'}}{\sqrt{(s-1-s')!}} |\xi^{s'}\rangle \,,
\nonumber\\
&& |\xi^{s'}\rangle = \frac{1}{s'!}  \alpha^{A_1} \ldots \alpha^{A_{s'}}  \xi^{A_1\ldots A_{s'}}|0\rangle\,, \hspace{2.1cm}  \hbox{ for massive field}.
\eeq

In terms of the ket-vectors $\phik$, $\xik$, gauge transformations  of the massless and massive fields can be presented on an equal footing as follows
\be \label{25062014-man-04x1}
\delta \phik   =   \Gbf \xik\,, \qquad  \Gbf \equiv \alphabf \Dbf - \ebf_1 - \alphabf^2 \frac{1}{2N_\alphabf +d-1}\bar\ebf_1\,,
\ee
where operators $\ebf_1$, $\bar\ebf_1$ are given in \rf{21062014-man-06x3}, \rf{21062014-man-06x4}, while $N_\alphabf$ is defined in Appendix.

\subsection{  BRST invariant Lagrangian and partition function }\label{secpartfun}

{\bf BRST invariant Lagrangian}.
To built BRST invariant Lagrangian one needs to introduce Faddeev-Popov and Nakanishi-Laudrup fields.%
\footnote{ In this paper, we deal with global BRST symmetries. Study of massless and massive fields with local BRST symmetries may be found in Refs.\cite{Buchbinder:2006ge}-\cite{Buchbinder:2013uha}.}
Generating form of Faddeev-Popov fields is described by  ket-vectors $\ck$, $\cbk$, while a generating form of Nakanishi-Laudrup fields is described by ket-vector $\bk$. Decomposition of these ket-vectors into scalar, vector, and tensor fields of  the $so(d,1)$ algebra takes the form
\beq
\label{26062014-man-06} && \hspace{-1cm} \ck = \frac{1}{(s-1)!}  \alpha^{A_1} \ldots \alpha^{A_{s-1}}  c^{A_1\ldots A_{s-1}}|0\rangle\,,  \hspace{1cm} \cbk = \frac{1}{(s-1)!}  \alpha^{A_1} \ldots \alpha^{A_{s-1}}  \cb^{A_1\ldots A_{s-1}}|0\rangle\,,
\nonumber\\
&& \hspace{-1cm} \bk = \frac{1}{(s-1)!}  \alpha^{A_1} \ldots \alpha^{A_{s-1}}  b^{A_1\ldots A_{s-1}}|0\rangle\,, \hspace{2.4cm} \hbox{ for massless field};
\eeq
\beq
\label{26062014-man-07} && \ck = \sum_{s'=0}^{s-1} \frac{\zeta^{s-1-s'}}{\sqrt{(s-1-s')!}} |c^{s'}\rangle \,, \qquad |c^{s'}\rangle = \frac{1}{s'!}  \alpha^{A_1} \ldots \alpha^{A_{s'}}  c^{A_1\ldots A_{s'}}|0\rangle\,,
\nonumber\\
&& \cbk = \sum_{s'=0}^{s-1} \frac{\zeta^{s-1-s'}}{\sqrt{(s-1-s')!}} |\cb^{s'}\rangle \,, \hspace{1cm}  |\cb^{s'}\rangle = \frac{1}{s'!}  \alpha^{A_1} \ldots \alpha^{A_{s'}}  \cb^{A_1\ldots A_{s'}}|0\rangle\,,
\nonumber\\
&& \bk = \sum_{s'=0}^{s-1} \frac{\zeta^{s-1-s'}}{\sqrt{(s-1-s')!}} |b^{s'}\rangle \,, \hspace{1cm} |b^{s'}\rangle = \frac{1}{s'!}  \alpha^{A_1} \ldots \alpha^{A_{s'}}  b^{A_1\ldots A_{s'}}|0\rangle\,,
\nonumber\\[-11pt]
&& \hspace{7.8cm}  \hbox{ for massive field}.
\eeq
Tensor fields \rf{26062014-man-06}, \rf{26062014-man-07} are totally symmetric traceless tensor fields of the $so(d,1)$ algebra.

Using the ket-vectors, the  BRST invariant Lagrangian $\LL_\tot$ in arbitrary  $\alpha$-gauge can be presented as
\beq
\label{25062014-man-17} && \LL_\tot = \LL + \LL_\qu \,,
\\
&& \frac{1}{e} \LL_\qu = - \langle b |\bar\Lbf\phik + \langle \cb| ( \Box_\pAdS + M_{_\FP} \bigr) \ck + \half \alpha \bbr\bk\,,
\\
&& M_{_\FP} \equiv - m^2  + \rho \Bigl( (s-1)(s+d-2) + N_\zeta(2 s + d - 3 -N_\zeta)  \Bigr) \,,
\eeq
where operator $\bar\Lbf$ is defined in \rf{21062014-man-06x1}.
Lagrangian \rf{25062014-man-17} is invariant under the following BRST and anti-BRST transformations (invariance of Lagrangian is assumed up to total derivative)

\beq
\label{25062014-man-18} &&  \ssf  \phik =   \Gbf \ck\,,  \qquad \ssf  \ck  =  0  \,,  \hspace{1.6cm} \ssf  |\cb\rangle  =    \bk \,, \qquad \ssf   \bk = 0 \,,\qquad
\\
\label{25062014-man-18x1} && \ssfb   \phik =   \Gbf |\cb\rangle\,,\qquad \ssfb   \ck  = -  \bk \,,\hspace{1cm}  \ssfb   |\cb\rangle  = 0\,, \hspace{1.2cm} \ssfb   \bk  =  0  \,,
\eeq
where $\Gbf$ is given in \rf{25062014-man-04x1}. BRST and anti-BRST transformations \rf{25062014-man-18},  \rf{25062014-man-18x1} are off-shell nilpotent: $\ssf^2=0$, $\ssfb^2=0$, $\ssf\ssfb+\ssfb\ssf=0$.

\noindent {\bf Partition function}. For the computation of a partition function, we chose the $\alpha=1$  gauge and integrate out Nakanishi-Laudrup fields. Doing so, we cast  Lagrangian \rf{25062014-man-17} into the form
\be \label{25062014-man-19}
\frac{1}{e} \LL_\tot = \half \phibr \mubf \bigl( \Box_{_\pAdS} + \mbf_1 +\rho\alphabf^2\bar\alphabf^2 \bigr)\phik  + \langle \cb| ( \Box_{_\pAdS} + M_{_\FP} \bigr) \ck\,.\qquad
\ee
We recall that a partition function does not depend on a choice of gauge condition. However we would like to emphasize that it is the representation for Lagrangian in  \rf{21062014-man-05}-\rf{21062014-man-06x2} and the $\alpha=1$ gauge that simplify considerably the expression for $\LL_\tot$ \rf{25062014-man-19} and hence the computation of a partition function.%
\footnote{ Recent research in Ref.\cite{Batalin:2014yxa} provides new interesting way for proving gauge independence of a partition function.}
Alternatively, Lagrangian \rf{25062014-man-19} can be represented as
\be \label{25062014-man-20}
\frac{1}{e} \LL_\tot =  \half  \phibr\Bigl( \Box_\pAdS + M_1 -  \frac{1}{4}\alphabf^2 \bar\alphabf^2 ( \Box_\pAdS + M_2) \Bigr)\phik +  \langle \cb| ( \Box + M_{_\FP} \bigr) \ck\,.
\ee
It is convenient to decompose the double-traceless ket-vector $\phik$ into two traceless ket-vectors $|\phi_{_\I} \rangle$, $|\phi_{_\II} \rangle$ by using the relations
\beq
\label{25062014-man-21} && \hspace{-1.5cm} \phik = |\phi_{_\I} \rangle   + \alphabf^2 \NN |\phi_{_\II} \rangle\,, \qquad  \bar\alphabf^2 |\phi_{_\I} \rangle =0\,,  \qquad \bar\alphabf^2 |\phi_{_\II} \rangle =0\,,
\\
&& \hspace{-1.5cm} \NN\equiv ((2s+d-3)(2s+d-5))^{-1/2}\,, \hspace{3cm} \hbox{ for massless field},
\\
&& \hspace{-1.5cm} \NN\equiv ((2s+d-3-2N_\zeta)(2s+d-5-2N_\zeta))^{-1/2}\,, \hspace{0.5cm} \hbox{ for massive field}. \qquad
\eeq
Plugging \rf{25062014-man-21} into \rf{25062014-man-20}, we obtain
\be \label{25062014-man-22}
\frac{1}{e} \LL_\tot =  \half \langle \phi_{_\I} | (\Box_{_{(A)dS}} + M_1) |\phi_{_\I} \rangle - \half \langle \phi_{_\II} | (\Box_{_{(A)dS}} + M_2 + 4\rho ) |\phi_{_\II} \rangle
+  \langle \cb |\bigl( \Box_{_{(A)dS}} + M_{_\FP} \bigl) |c\rangle\,.
\ee
Decomposition of ket-vectors $|\phi_{_\I} \rangle$,  $|\phi_{_\II} \rangle$ \rf{25062014-man-21} into scalar, vector, and traceless tensor fields of the $so(d,1)$ algebra is given by
\beq
&& |\phi_\I\rangle = \frac{1}{s!}  \alpha^{A_1} \ldots \alpha^{A_s}  \phi_\I^{A_1\ldots A_s}|0\rangle\,,
\nonumber\\
\label{21062014-man-03y1} && |\phi_\II\rangle = \frac{1}{(s-2)!}  \alpha^{A_1} \ldots \alpha^{A_{s-2}}  \phi_\II^{A_1\ldots A_{s-2}}|0\rangle\,,\hspace{1.5cm} \hbox{ for massless field};\qquad
\\
&& |\phi_\I\rangle  = \sum_{s'=0}^{s} \frac{\zeta^{s-s'}}{\sqrt{(s-s')!}} |\phi_\I^{s'}\rangle \,, \qquad |\phi_\II\rangle  = \sum_{s'=0}^{s-2} \frac{\zeta^{s-2-s'}}{\sqrt{(s-2-s')!}} |\phi_\II^{s'}\rangle \,,
\nonumber\\
&& |\phi_{\I,\II}^{s'}\rangle = \frac{1}{s'!}  \alpha^{A_1} \ldots \alpha^{A_{s'}}  \phi_{\I,\II}^{A_1\ldots A_{s'}}|0\rangle\,, \hspace{2.8cm}  \hbox{ for massive field}.
\eeq
In terms of the scalar, vector, and tensor fields, Lagrangian \rf{25062014-man-22}  can be presented as
\beq
\label{26062014-man-14} &&   \hspace{-1cm}  \LL_\tot =    \LL_\I^s -   \LL_\II^{s-2} +     \LL_\FP^{s-1}\,, \hspace{5.4cm} \hbox{ for massless field},
\\
&&   \hspace{-1cm} \LL_\tot =  \sum_{s'= 0}^s \LL_\I^{s'} -  \sum_{s'=0}^{s-2} \LL_\II^{s'} + \sum_{s'= 0}^{s-1}  \LL_\FP^{s'}\,, \hspace{3.6cm} \hbox{ for massive field},\qquad
\\
\label{26062014-man-15} &&  \hspace{-1cm}    \LL_{\I,\II}^{s'}  = \frac{e}{2s'!}\, \phi_{\I,\II}^{A_1 \ldots A_{s'}} (\DD^2 + M_{1}^{s'})  \phi_{\I,\II}^{A_1 \ldots A_{s'}}\,, \qquad \LL_\FP^{s'}  = \frac{e}{s'!}\, \cb^{A_1 \ldots A_{s'}} (\DD^2 + M_1^{s'})  c^{A_1 \ldots A_{s'}}\,,
\\
&& \hspace{-1cm} M_1^s \equiv \rho \bigl((s-2)(s+d-2) - s \bigr) \,, \hspace{1cm}  \qquad M_1^{s-1} \equiv \rho (s-1)(s+d-2)  \,,
\nonumber\\
&& \hspace{-1cm} M_1^{s-2} \equiv \rho \bigl( s (s+d-1) -2 \bigr) \,, \hspace{5cm} \hbox{ for massless field},\qquad
\\
\label{26062014-man-17} &&  \hspace{-1cm} M_1^{s'} \equiv -m^2 + \rho \bigl(2(s-1)(s+d-2) - s'(s'+d-1)\bigr)\,, \hspace{0.5cm} \hbox{ for massive field}.\qquad
\eeq
From \rf{26062014-man-14}-\rf{26062014-man-17}, we see that the partition function is given by
\beq
\label{26062014-man-03} Z & = &  D^{s-1} D^{s-1}  \Bigr/  D^s  D^{s-2}\,, \hspace{3cm} \hbox{ for massless field},
\\
\label{26062014-man-04}   Z & = & \prod_{s'=0}^{s-1}  D^{s'} D^{s'}  \Bigr/ \prod_{s'=0}^s  D^{s'}   \prod_{s'=0}^{s-2}  D^{s'}
\nonumber\\
& = &  D^{s-1} \bigr/ D^s, \hspace{ 5cm} \hbox{ for massive field},\qquad
\\
\label{26062014-man-05} &&   D^{s'} \equiv \bigl(\det (-\DD^2 - M_1^{s'}) \bigr)^{1/2}\,,
\eeq
where, in relation \rf{26062014-man-05}, the determinant of D'Alembert operator is evaluated for rank-$s'$ traceless tensor field. Often, it is convenient to use the following relation for $D^{s'} $ \rf{26062014-man-05}:
\be \label{26062014-man-09}
D^{s'} = D^{s'\perp} D^{s'-1}\,,
\ee
where $D^{s'\perp}$ takes the same form as in \rf{26062014-man-05}, while the determinant of D'Alembert operator is evaluated on space of traceless and divergence-free rank-$s'$ tensor field.  Using \rf{26062014-man-09} in \rf{26062014-man-03},\rf{26062014-man-04}, we obtain
\beq
\label{26062014-man-10} && Z =  D^{s-1\perp}\bigr/ D^{s\perp} \hspace{2.5cm} \hbox{ for massless field},\qquad
\\
\label{26062014-man-11} && Z = 1\bigr/D^{s\perp} \hspace{3.4cm} \hbox{ for massive field}.\qquad
\eeq
In earlier literature, partition functions \rf{26062014-man-10}, \rf{26062014-man-11} were obtained by different methods in Refs.\cite{Gupta:2012he,Tseytlin:2013jya}.

\newsection{  \large $so(d-1,1)$ covariant approach and effective action  }

In this section, we review Lagrangian $so(d-1,1)$ covariant approach to dynamics of free fields in  AdS${}_{d+1}$ developed in Refs.\cite{Metsaev:2008ks,Metsaev:2009hp}.  In Sec.\ref{sec brst-2}, we use our approach for the derivation of BRST invariant Lagrangians of AdS fields and apply such Lagrangians for derivation of BRST invariant effective action of shadow fields.

To develop  $so(d-1,1)$ covariant approach to fields in AdS$_{d+1}$ we use the Poincar\'e parametrization of AdS$_{d+1}$,
\be \label{23062014-man-01}
ds^2 = \frac{1}{z^2}(dx^a dx^a + dz dz)\,.
\ee
From \rf{23062014-man-01}, we see that the line element respects manifest symmetries of the $so(d-1,1)$ algebra.

\noindent {\bf Field content}. In the framework of metric-like approach,  for the gauge invariant description of spin-$s$ massless field in  AdS$_{d+1}$, we use the following scalar, vector and totally symmetric double-traceless tensor field of the $so(d-1,1)$ algebra,
\be \label{23062014-man-02}
\phi^{a_1\ldots a_{s'}}\,, \hspace{1cm} s'=0,1,\ldots,s\,, \qquad \phi^{aabba_5\ldots a_{s'}}=0\,, \qquad  \hbox{ for } s' \geq 4\,,
\ee
while, for the gauge invariant description of spin-$s$ massive field in  AdS$_{d+1}$, we use the following scalar, vector and totally symmetric double-traceless tensor fields of the  $so(d-1,1)$ algebra:
\be  \label{23062014-man-03}
\phi_\lambda^{a_1\ldots a_{s'}}\,,
\qquad
s'=0,1,\ldots, s\,,
\qquad
\lambda \in [s-s']_2\,, \qquad \phi_\lambda^{aabba_5\ldots a_{s'}}=0\,, \quad  \hbox{ for } s' \geq 4\,.
\ee
Here and below the notation $\lambda \in [n]_2$ implies that $\lambda$ takes the following values: $\lambda =-n,-n+2,-n+4,\ldots,n-4, n-2,n$.

To simplify the presentation of gauge gauge-invariant Lagrangian we use generating form of gauge fields. To this end we introduce the oscillators $\alpha^a$, $\alpha^z$, $\zeta$ and note that gauge fields in \rf{23062014-man-02}, \rf{23062014-man-03} can be collected into a ket-vector $\phik$  as  follows
\beq
\label{23062014-man-04} &&  \hspace{-2.5cm} |\phi\rangle \equiv \sum_{s'=0}^s
\frac{\alpha_z^{s-s'}}{\sqrt{(s - s')!}}|\phi^{s'}\rangle \,,
\nonumber\\
&&  \hspace{-2.5cm} |\phi^{s'}\rangle \equiv \frac{1}{s'!} \alpha^{a_1} \ldots \alpha^{a_{s'}}
\, \phi^{a_1\ldots a_{s'}} |0\rangle\,,  \hspace{4cm} \hbox{ for massless field},
\\
\label{23062014-man-05} && \hspace{-2.5cm}  |\phi\rangle = \sum_{s'=0}^s |\phi^{s'}\rangle \,,
\nonumber\\
&& \hspace{-2.5cm} |\phi^{s'}\rangle   =  \sum_{\lambda \in [s-s']_2}\!\!
\frac{\zeta_{\phantom{z}}^{\frac{s-s'+\lambda}{2}}
\alpha_z^{\frac{s-s'-\lambda}{2}}\alpha^{a_1}\ldots
\alpha^{a_{s'}}}{s'!\sqrt{(\frac{s-s'+\lambda}{2})!
(\frac{s-s'-\lambda}{2})!}} \, \phi_\lambda^{a_1\ldots a_{s'}} |0\rangle\,, \qquad \hbox{ for massive field}.
\eeq

{\bf Gauge invariant Lagrangian}. In terms of ket-vectors $\phik$ \rf{23062014-man-04},\rf{23062014-man-05}, gauge invariant action and Lagrangian for massless and massive fields can be presented on an equal footing as follows
\beq \label{23062014-man-06}
&&  S   =  \int  d^dx\,  dz\,   \LL \,,
\\
\label{23062014-man-07} && \LL  =    \half  \langle  \phi|\mu (\Box - \MM^2) |\phi\rangle  +  \half  \langle  \Lb\phi| |\Lb \phi \rangle\,,
\\
\label{23062014-man-08} && \hspace{1.3cm}   \Lb  \equiv
\bar\alpha \partial  -  \half \alpha \partial  \bar\alpha^2 -
\eb_1 \Pi^\smponetwo  +  \half e_1  \bar\alpha^2\,,
\eeq
$\Box \equiv \partial^a\partial^a$, $\partial^a = \eta^{ab}\partial /\partial x^b$, $| \Lb \phi\rangle \equiv   \Lb |\phi\rangle$. Expressions for scalar products like  $\alpha\partial$, $\alpha^2$, $ \mu$, $\Pi^\smponetwo$ are defined in Appendix. Bra-vectors $\langle\phi |$, $\langle  \Lb  \phi |$ are defined as follows,  $\langle\phi | \equiv (|\phi \rangle)^\dagger$, $\langle  \Lb  \phi | \equiv (| \Lb  \phi \rangle)^\dagger$. Operators $\MM^2$, $e_1$, $\eb_1$ are given in Table.
From \rf{23062014-man-07}, we see that the kinetic terms of massless and massive fields have one and the same dependence on the vector oscillators $\alpha^a$ and the derivatives $\partial^a$. From Table, we see that that all dependence of the kinetic terms of the massless and massive fields on the scalar oscillators  $\zeta$, $\alpha^z$, the radial coordinate $z$, and the radial derivative $\partial_z =\partial/\partial z$ is described completely by the operators $\MM^2$, $e_1$, $\eb_1$.

\noindent {\bf Gauge symmetries}. For the discussion of gauge symmetries, we introduce gauge transformation parameters. Generating form of gauge transformation parameters is described by a ket-vector $\xik$.
Decomposition of the ket-vector into scalar, vector, and totally symmetric traceless tensor fields of the $so(d-1,1)$ algebra is given by
\beq
\label{27062014-man-01} && \hspace{-1.7cm} |\xi\rangle \equiv \sum_{s'=0}^{s-1}
\frac{\alpha_z^{s-1-s'}}{\sqrt{(s -1 - s')!}} |\xi^{s'}\rangle \,,
\nonumber\\
&& \hspace{-1.7cm} |\xi^{s'}\rangle \equiv
\frac{1}{s'!} \alpha^{a_1} \ldots \alpha^{a_{s'}} \xi^{a_1\ldots a_{s'}} |0\rangle\,,  \hspace{5.1cm} \hbox{ for massless field},
\\
\label{27062014-man-02} && \hspace{-1.7cm} |\xi\rangle  = \sum_{s'=0}^{s-1} |\xi^{s'}\rangle\,,
\nonumber\\
&& \hspace{-1.7cm} |\xi^{s'}\rangle = \sum_{\lambda \in [s-1-s']_2}\!\!\!
\frac{\zeta_{\phantom{z}}^{\frac{s-1-s'+\lambda}{2}}
\alpha_z^{\frac{s-1-s'-\lambda}{2}}\alpha^{a_1}\ldots
\alpha^{a_{s'}}}{s'!\sqrt{(\frac{s-1-s'+\lambda}{2})!
(\frac{s-1-s'-\lambda}{2})!}} \, \xi_\lambda^{a_1\ldots a_{s'}} |0\rangle\,, \qquad \hbox{for massive field}.
\eeq

In terms of the ket-vectors $\phik$, $\xik$ above discussed, gauge transformations of massless and massive fields can be presented on an equal footing,
\be \label{27062014-man-03}
\delta \phik   =   G \xik\,, \qquad  G \equiv \alpha  \partial  - e_1 - \alpha^2 \frac{1}{2N_\alpha +d-2}\eb_1\,,
\ee
where the operators $e_1$, $\eb_1$ are given in the Table, while $N_\alpha$ is  defined in Appendix.

\newpage

\noindent {\sf Table. Operators $\MM^2$, $e_1$, $\eb_1$ entering  Lagrangian, gauge transformations and BRST transformations of AdS field in the framework of $so(d-1,1)$ covariant formulation.  $m$ is mass parameter of massive field. In Table, we present also the operators $e_1$, $\eb_1$ entering BRST transformations canonical and anomalous shadows.}

\begin{center}
\begin{tabular}{|l|c|c|c|}
\hline &&&
\\[-3mm]
Fields   & $\MM^2$  & $e_1$  &  $\eb_1$
\\ [1mm]\hline
&&&
\\[-1mm]
massless spin-$s$&  $ - \partial_z^2 + \frac{1}{z^2} (\nu^2-\frac{1}{4})$ & $ - \alpha^z e_z \TT_{\nu-\half}$ & $ -  \TT_{-\nu + \half} e_z \bar\alpha^z$
\\[3mm] \cline{2-4}\\[-5mm]
field in AdS${}_{d+1}$ & \mc{3}{|c|}{} \\
& \mc{3}{|c|}{ $\nu \equiv s+\frac{d-4}{2} - N_z$, \qquad $\TT_\nu\equiv \partial_z +\frac{\nu}{z}$ \qquad $\partial_z\equiv \partial/\partial z$}
\\[2mm]\hline
&  &   &
\\[0mm]
massive spin-$s$ & $- \partial_z^2 + \frac{1}{z^2} (\nu^2-\frac{1}{4})$ & $ - \zeta r_\zeta \TT_{ -\nu - \half} - \alpha^z r_z \TT_{\nu-\half}$ & $
- \TT_{\nu + \half}  r_\zeta \bar\zeta  - \TT_{-\nu + \half} r_z
\bar\alpha^z $
\\[3mm] \cline{2-4}\\[-5mm]
field in AdS${}_{d+1}$ & \mc{3}{|c|}{} \\
& \mc{3}{|c|}{ $\nu \equiv \kappa + N_\zeta - N_z$, \qquad $\TT_\nu\equiv \partial_z +\frac{\nu}{z}$\,, \qquad $\partial_z\equiv \partial/\partial z$  }
\\[2mm]\hline
&&&
\\[-3mm]
canonical spin-$s$  & - & $ \alpha^z e_z \Box $  & $ - e_z \bar\alpha^z  $
\\[1mm]
shadow in $R^{d-1,1}$ &  &   &
\\[-1mm]\hline
 &   &   &
\\[-2mm]
anomalous spin-$s$  & - & $ \zeta r_\zeta  + \alpha^zr_z \Box $  & $ -r_\zeta\bar\zeta \Box - r_z \bar\alpha^z  $
\\[1mm]
shadow in $R^{d-1,1}$ &  &   &
\\[0mm]\hline
\mc{4}{|c|}{}
\\
\mc{4}{|c|}{ $e_z = \Bigl(\frac{2s+d-4-N_z}{2s+d-4-2N_z}\Bigr)^{1/2}$}
\\
\mc{4}{|c|}{}
\\
\mc{4}{|c|}{ $ r_\zeta = \left(\frac{(s+\frac{d-4}{2}
-N_\zeta)(\kappa - s-\frac{d-4}{2} + N_\zeta)(\kappa + 1 +
N_\zeta)}{2(s+\frac{d-4}{2}-N_\zeta - N_z)(\kappa +N_\zeta -N_z) (\kappa+
N_\zeta - N_z +1)}\right)^{1/2} $ }
\\
\mc{4}{|c|}{}
\\
\mc{4}{|c|}{ $ r_z = \left(\frac{(s+\frac{d-4}{2} -N_z)(\kappa
+ s + \frac{d-4}{2} - N_z)(\kappa - 1 -
N_z)}{2(s+\frac{d-4}{2}-N_\zeta-N_z)(\kappa + N_\zeta - N_z) (\kappa +N_\zeta
- N_z -1)}\right)^{1/2} $ }
\\
\mc{4}{|c|}{}
\\
\mc{4}{|c|}{ $ \kappa \equiv \sqrt{m^2 + \Bigl( s+
\frac{d-4}{2}\Bigr)^2} $ }
\\ [3mm]\hline
\end{tabular}
\end{center}

\subsection{\large  BRST invariant Lagrangian of AdS fields and effective action of shadow fields }\label{sec brst-2}

{\bf BRST invariant Lagrangian of AdS field}. To built BRST invariant Lagrangian we should introduce Faddeev-Popov and Nakanishi-Laudrup fields. Generating form of Faddeev-Popov fields is described by  ket-vectors $\ck$, $\cbk$, while generating form of Nakanishi-Laudrup fields is described by ket-vector $\bk$. The decomposition of these ket-vectors into the corresponding scalar, vector, and traceless tensor fields $\cb^{a_1\ldots a_{s'}}$, $c^{a_1\ldots a_{s'}}$, $b^{a_1\ldots a_{s'}}$  takes the same form as the one in \rf{27062014-man-01}, \rf{27062014-man-02} and can be obtained by using the following substitutions in  \rf{27062014-man-01}, \rf{27062014-man-02},
\be
\xi \rightarrow c\,, \qquad
\xi \rightarrow \cb\,, \qquad
\xi \rightarrow b \,.
\ee
By definition, tensorial fields entering the ket-vectors $\ck$, $\cbk$, $\bk$  are totally symmetric traceless tensor fields of the $so(d-1,1)$ algebra.

In terms of the ket-vectors, BRST invariant Lagrangian $\LL_\tot$ in arbitrary  $\alpha$-gauge can be presented as
\be
\label{27062014-man-04}  \LL_\tot = \LL + \LL_\qu \,, \qquad \LL_\qu = - \langle b |\Lb\phik + \langle \cb| ( \Box - \MM^2\bigr) \ck + \half \alpha \bbr\bk\,,
\ee
where operator $\Lb$ is given in \rf{23062014-man-08}, while the operator $\MM^2$ is defined in Table. Lagrangian \rf{27062014-man-04} is invariant under the following BRST and anti-BRST transformations:
\beq
\label{27062014-man-05} && \ssf  \phik =   G \ck\,,  \qquad \ssf  \ck  =  0  \,,  \hspace{1.6cm} \ssf  |\cb\rangle  =    \bk \,, \qquad \ssf   \bk = 0 \,,\qquad
\\
\label{27062014-man-05x1} && \ssfb   \phik =   G |\cb\rangle\,,\qquad \ssfb   \ck  = -  \bk \,,\hspace{1cm} \ssfb   |\cb\rangle  = 0\,, \hspace{1.2cm} \ssfb   \bk  =  0  \,,
\eeq
where the  operator $G$ is given in \rf{27062014-man-03}.
BRST and anti-BRST transformations \rf{27062014-man-05}, \rf{27062014-man-05x1} are off-shell nilpotent: $\ssf^2=0$, $\ssfb^2=0$, $\ssf\ssfb+\ssfb\ssf=0$.

For the computation of effective action of shadow fields, we choose the $\alpha=1$ gauge and integrate out the Nakanishi-Laudrup fields. Doing so, we cast the BRST invariant Lagrangian \rf{27062014-man-04} into the form
\beq
\label{27062014-man-06} && \LL_\tot  =    \half  \langle  \phi|\mu (\Box - \MM^2) |\phi\rangle + \cbbr (\Box - \MM^2) \ck.
\eeq
BRST and anti-BRST symmetries of Lagrangian \rf{27062014-man-06} are realized by the following transformations:
\beq
\label{27062014-man-06x1} && \ssf  \phik  =     G  \ck\,, \qquad \ssf  \ck = 0\,, \hspace{1.7cm}  \ssf  \cbk   =  \Lb  \phik \,,
\\
\label{27062014-man-06x2} && \ssfb   \phik  =    G  \cbk\,,\qquad\ssfb   \ck = -   \Lb \phik\,,   \qquad \ssfb   \cbk   = 0 \,,
\eeq
where the operators $\Lb$ and $G$ are given in \rf{23062014-man-08} and \rf{27062014-man-03} respectively. Transformations  \rf{27062014-man-06x1}, \rf{27062014-man-06x2} are nilpotent only for on-shell Faddeev-Popov fields.

\noindent {\bf AdS/CFT correspondence}.  AdS/CFT correspondence is realized in two steps, at least. First, we solve equations of motion for AdS field with a suitable boundary conditions, i.e., we solve the Dirichlet problem. Boundary conditions are fixed by requiring the boundary value of AdS field  to be related to some particular representation of conformal algebra which is referred to as shadow field.  Namely the boundary values of massless AdS field and massive AdS field correspond to the representations of the conformal algebra which are referred to as canonical shadow field and anomalous shadow field respectively.  Second, we plug solution of the Dirichlet problem into action of AdS field. Action of AdS field evaluated on the solution of the Dirichlet problem is referred to as effective action. For massless field, the effective action is functional of canonical shadow, while for massive field, the effective action is functional of anomalous shadow. We recall that, for free AdS field, kernel of effective action is a 2-point correlation function of CFT. Note that in our approach we solve the Dirichlet problem not only for gauge fields but also for Faddeev-Popov fields, i.e., we introduce boundary Faddeev-Popov shadow fields. This leads to BRST invariant effective action.  We now describe details of the computation of the BRST invariant effective action.

Equations of motion for gauge fields and Faddeev-Popov fields obtained form Lagrangian \rf{27062014-man-06} take the form
\be
\label{27062014-man-07}   \Box_\nu \phik = 0 \,, \qquad \Box_\nu \cbk = 0 \,, \qquad \Box_\nu \ck = 0\,, \qquad
\Box_\nu \equiv \Box + \partial_z^2 - \frac{1}{z^2}(\nu^2 -\frac{1}{4})\,.\qquad %
\ee
Solution to the Dirichlet problem for equations of motion \rf{27062014-man-07}  with boundary conditions for the gauge field $\phik$ and the Faddeev-Popov fields $\ck$, $\cbk$ corresponding to the respective shadow gauge fields, denoted by $|\phi_\sh\rangle$, and shadow Faddeev-Popov fields, denoted by $|c_\sh\rangle$, $|\cb_\sh\rangle$, can be presented as
\beq
\label{27062014-man-08} |\phi(x,z)\rangle  & = &  \sigma_\nu \int d^dy\, G_\nu (x-y,z)
|\phi_\sh(y)\rangle\,,
\\
\label{27062014-man-09} |c(x,z)\rangle  & = &  \sigma_\nu \int d^dy\, G_\nu (x-y,z)
|c_\sh(y)\rangle\,,
\\
\label{27062014-man-10} |\cb(x,z)\rangle  & = &  \sigma_\nu \int d^dy\, G_\nu (x-y,z)
|\cb_\sh(y)\rangle\,,
\\
\label{27062014-man-11} && \sigma_\nu \equiv \frac{2^\nu\Gamma(\nu)}{
2^\kappab\Gamma(\kappab)}(-)^{N_z} \,,
\eeq
where the Green function $G_\nu$ is given by
\beq
\label{27062014-man-12} && G_\nu(x,z) = \frac{c_\nu z^{\nu+\half}}{ ( |x|^2 + z^2)^{\nu + \frac{d}{2}} }\,,   \qquad  c_\nu \equiv \frac{\Gamma(\nu+\frac{d}{2})}{\pi^{d/2} \Gamma(\nu)} \,,
\eeq
Note that, for massless fields, relations in \rf{27062014-man-08}-\rf{27062014-man-10} provide solution of the Dirichlet problem with the canonical shadows $|\phi_\sh\rangle$, $|c_\sh\rangle$, $|\cb_\sh\rangle$ as boundary data, while, for massive fields, relations in \rf{27062014-man-08}-\rf{27062014-man-10} provide solution of the Dirichlet problem with the anomalous shadows  $|\phi_\sh\rangle$, $|c_\sh\rangle$, $|\cb_\sh\rangle$ as boundary data.  Also note that, the decomposition of $|\phi_\sh\rangle$ into scalar, vector, and double-traceless tensor fields of the $so(d-1,1)$ algebra takes the same form as in \rf{23062014-man-04},\rf{23062014-man-05},
while the decomposition of $|c_\sh\rangle$,  $|\cb_\sh\rangle$,  into scalar, vector, and traceless tensor fields of the $so(d-1,1)$ algebra takes the same form as in \rf{27062014-man-01},\rf{27062014-man-02}. The $\nu$ and $\bar\kappa$ appearing in \rf{27062014-man-08}-\rf{27062014-man-12} are given by
\beq
\label{28062014-man-01} &&\hspace{-0.5cm}  \nu \equiv s+ \frac{d-4}{2} - N_z, \qquad \bar\kappa \equiv  s + \frac{d-4}{2}, \hspace{1cm} \hbox{ for canonical shadows},\qquad
\\
&& \hspace{-0.5cm} \nu \equiv \kappa + N_\zeta -N_z, \hspace{1.4cm} \bar\kappa \equiv \kappa, \hspace{2.4cm} \hbox{ for anomalous shadows},
\eeq
where $\kappa$ is given in Table.

Taking into account \rf{27062014-man-08}-\rf{27062014-man-10},
and the asymptotic behavior of the Green function
\be \label{27062014-man-14}
G_\nu(x,z) \ \ \ \stackrel{z \rightarrow 0}{\longrightarrow} \ \ \ z^{-\nu
+ \half} \delta^d(x)\,,
\ee
we get the asymptotic behavior of solution of the Dirichlet problem for the gauge fields
\be \label{27062014-man-15}
|\phi(x,z)\rangle  \,\,\, \stackrel{z\rightarrow 0 }{\longrightarrow}\,\,\,
z^{-\nu + \half} \sigma_\nu |\phi_\sh(x)\rangle
\ee
and similar asymptotic behavior for the solution of Faddeev-Popov fields.

To find the effective action, we should plug our solution \rf{27062014-man-08}-\rf{27062014-man-10} into action \rf{23062014-man-06} with Lagrangian \rf{27062014-man-06}. Note also that we should add to the action an appropriate boundary term. Using general method for finding boundary term in Ref.\cite{Arutyunov:1998ve}, we make sure that Lagrangian which involves a contribution of the boundary term can be presented as
\be \label{27062014-man-16}
\LL_\tot  =   \half \langle \partial^a \phi|\mu  | \partial^a \phi\rangle +\half
\langle \TT_{\nu-\half} \phi| \mu   | \TT_{\nu-\half} \phi\rangle
+ \langle \partial^a \cb|   | \partial^a c \rangle +
\langle \TT_{\nu-\half} \cb |     | \TT_{\nu-\half} c\rangle\,,
\ee
where $\TT_\nu$ is defined in the Table. Note that, in order to adapt our formulas to Euclidean signature, we change sign of Lagrangian, $\LL\rightarrow - \LL$, when passing from \rf{27062014-man-06} to \rf{27062014-man-16}.
It is easy to verify that action \rf{23062014-man-06}, \rf{27062014-man-16} considered on the solution of equations of motion
can be represented as%
\footnote{ For the discussion of AdS/CFT correspondence, we use a Lagrangian approach. The study of AdS/CFT correspondence by using the higher-spin symmetries may be found in Refs.\cite{Didenko:2012tv}.}
\be
\label{27062014-man-17} - S_\eff^\tot  =  \int d^dx\,  \LL_\eff^\tot\Bigr|_{z\rightarrow
0} \,, \qquad  \ \ \
\LL_\eff^\tot  =  \half \phibr \mu \TT_{\nu -\half } \phik +  \cbbr \TT_{\nu -\half } \ck \,.
\ee

\noindent {\bf BRST invariant effective action of shadow field}. Plugging \rf{27062014-man-08}-\rf{27062014-man-10} into \rf{27062014-man-17}, we get the following effective action:
\be  \label{27062014-man-18}
-S_\eff^\tot  =  2\kappab c_\kappab \Gamma_\tot \,,
\ee
where $\Gamma_\tot$ is given by
\beq
\label{25062014-man-39} && \Gamma_\tot \equiv \int d^dx_1 d^dx_2 \LL_{12}^\tot\,,
\\
\label{25062014-man-40} && \hspace{1.3cm} \LL_{12}^\tot \equiv \half \langle\phi_\sh(x_1)|
\frac{\mu f_\nu}{ |x_{12}|^{2\nu + d }} |\phi_\sh (x_2)\rangle
+  \langle \cb_\sh(x_1) | \frac{f_\nu}{ |x_{12}|^{2\nu + d }} | c_\sh (x_2)\rangle \,,
\\
\label{25062014-man-41} && \hspace{1.3cm} f_\nu \equiv \frac{\Gamma(\nu + \frac{d}{2})\Gamma(\nu + 1)}{4^{\bar\kappa - \nu}
\Gamma(\bar\kappa + \frac{d}{2})\Gamma(\bar\kappa + 1)} \,,
\\
&& \hspace{1.3cm} |x_{12}|^2 \equiv x_{12}^a x_{12}^a\,, \qquad x_{12}^a = x_1^a - x_2^a\,,
\eeq
and the operators $N_\zeta$, $N_z$, $\mu$ are defined in Appendix. Recall that  $c_\kappa$ is given in \rf{27062014-man-12},  while the parameter $\kappa$ is defined in Table.

Effective action \rf{27062014-man-18}  is invariant under the following BRST and anti-BRST transformations:
\beq
\label{28062014-man-02} &&  \ssf  |\phi_\sh\rangle  =     G  |c_\sh\rangle \,, \qquad  \ssf  |c_\sh\rangle  =0\,, \hspace{2.1cm}  \ssf  | \cb_\sh \rangle   =    \Lb  |\phi_\sh\rangle \,,\qquad
\\
\label{28062014-man-03} &&  \ssfb   |\phi_\sh\rangle  =     G  |\cb_\sh\rangle \,,\qquad \ssfb   |c_\sh\rangle  = - \Lb  |\phi_\sh\rangle \,, \qquad \ssfb   | \cb_\sh \rangle   = 0\,,
\\
\label{28062014-man-04}  && \qquad G \equiv   \alpar - e_1    -  \alpha^2\frac{1}{2N_\alpha +
d- 2} \eb_1 \,,
\\
\label{28062014-man-05}  && \qquad \Lb  \equiv  \albpar - \half \alpar \bar\alpha^2  - \eb_1 \Pi^\smponetwo + \half
e_1 \bar\alpha^2  \,,
\eeq
where operators $e_1$, $\eb_1$ corresponding to the canonical and anomalous shadows are given in Table.
BRST and anti-BRST transformations given in \rf{28062014-man-02}, \rf{28062014-man-03} are nilpotent.

\newsection{ \large BRST invariant Lagrangian of conformal fields }

For canonical shadows, a kernel of effective action \rf{25062014-man-39} is not well defined when $d$ is even integer (see, e.g.,
\cite{Aref'eva:1998nn}). Hopefully, the kernel becomes well defined upon using a dimensional regularization. When removing the regularization, we are left with a logarithmic divergence of  the kernel. Below, we demonstrate that the logarithmic divergence of the BRST invariant effective action turns out to be BRST invariant action of a conformal field.

Using the notation $[d]$ for integer part of $d$, we introduce the regularization parameter $\varepsilon$ by the relation
\be \label{28062014-man-09}
 d- [d]= - 2\varepsilon\,,\qquad [d]-\hbox{ even integer}.
\ee
Using \rf{28062014-man-09} and taking into account the dependence of $\nu$ on $d$ in\rf{28062014-man-01}, we note the following textbook asymptotic behavior for the kernel:
\be \label{25062014-man-05}
\frac{1}{|x|^{2\nu+d}}\,\,\, \stackrel{\varepsilon \sim
0}{\mbox{\Large$\sim$}}\,\,\, \frac{1}{\varepsilon} \varrho_\nu \Box^\nu
\delta^{(d)}(x)\,, \qquad   \varrho_\nu \equiv  \frac{\pi^{d/2}}{4^\nu \Gamma(\nu
+ 1)\Gamma(\nu + \frac{d}{2})}\,.
\ee
Plugging \rf{25062014-man-05} into expression for $\Gamma_\tot$ in \rf{25062014-man-39}, we obtain
\beq
\label{25062014-man-06} && \Gamma_\tot \,\,\, \stackrel{ \varepsilon \sim
0}{\mbox{\Large$\sim$}}\,\,\, \frac{1}{\varepsilon} \varrho_{\nu_s} \int
d^dx\,\, \LL_\tot\,, \hspace{1cm} \hbox{ for canonical shadows},\qquad
\nonumber\\
\label{25062014-man-07} && \LL_\tot = \half \phibr \mu  \Box^\nu \phik +  \langle \cb | \Box^\nu \ck\,, \qquad \nu \equiv s+ \frac{d-4}{2}-N_z \,.
\eeq
Note that, in order to simplify the notation, we make  the  identifications of the ket-vectors, $\phik\equiv |\phi_\sh\rangle$, $\ck\equiv |c_\sh\rangle$, $\cbk\equiv |\cb_\sh\rangle$, when passing from \rf{25062014-man-40} to \rf{25062014-man-06}. Lagrangian \rf{25062014-man-06} is BRST invariant Lagrangian of spin-$s$ conformal field.%
\footnote{ In this paper, we deal with global BRST transformations of free conformal fields. A study of free conformal fields with local BRST symmetries may be found in Refs.\cite{Bekaert:2013zya}. Discussion of general structure of interacting conformal fields in $3d$ may be found in Ref.\cite{Vasiliev:2012vf}.}
For the illustration purposes, we consider the Lagrangian for spin-1, spin-2, and arbitrary spin-$s$ fields in turn.

\subsection{  Spin-1 conformal field}

For spin-1 conformal field, Lagrangian \rf{25062014-man-07} takes the form
\be \label{25062014-man-08}
\LL_\tot = \half \phi^a \Box^{k+1}\phi^a  + \half \phi
\Box^k \phi + \cb \Box^{k+1}  c \,, \qquad k \equiv \frac{d-4}{2}\,.
\ee
Lagrangian \rf{25062014-man-08} is invariant under BRST and anti-BRST transformations given by
\beq
\label{25062014-man-09}   && \ssf \phi^a =   \partial^a c\,, \qquad \ssf \phi=  - \Box c \,,
\qquad \ssf c = 0 \,, \hspace{2.5cm} \ssf \cb=      \partial^a \phi^a + \phi \,, \qquad %
\\
\label{25062014-man-09x1}  &&  \ssfb \phi^a =   \partial^a \cb\,, \qquad \ssfb \phi=  - \Box \cb \,,
\qquad \ssfb c = -     \partial^a \phi^a - \phi\,, \qquad \ssfb \cb=  0\,.
\eeq
It is easy to verify that transformations \rf{25062014-man-09},\rf{25062014-man-09x1}  are off-shell nilpotent. This is related to the fact that the scalar field $\phi$ can be realized as the Nakanishi-Laudrup field. To see this we introduce a new field $b$ by the following relation
\be \label{25062014-man-11}
b = \phi + \partial^a\phi^a.
\ee
Plugging  $\phi = b - \partial^a\phi^a$ into \rf{25062014-man-08} we cast  Lagrangian \rf{25062014-man-08} into the form
\beq
\label{25062014-man-14} && \LL_\tot = \LL + \LL_\qu\,,
\\
\label{25062014-man-15} && \LL = - \frac{1}{4} F^{ab}\Box^k F^{ab}\,, \qquad F^{ab}\equiv \partial^a \phi^b - \partial^b \phi^a\,,
\\
\label{25062014-man-16} && \LL_\qu = - b \Box^k \partial^a\phi^a + \half b \Box^k b + \cb \Box^{k+1} c\,,
\eeq
while BRST and anti-BRST transformations \rf{25062014-man-09}, \rf{25062014-man-09x1} can be cast into the form
\beq
\label{25062014-man-09x2} && \ssf \phi^a =   \partial^a c\,, \qquad \ssf b= 0 \,,
\qquad \ssf c = 0\,,   \hspace{1.1cm} \ssf \cb=     b \,,
\\
\label{25062014-man-09x3} && \ssfb \phi^a =   \partial^a \cb\,, \qquad \ssfb b= 0 \,,
\qquad \ssfb c = -b \,, \qquad \ssfb \cb=    0 \,.
\eeq

From \rf{25062014-man-16}, \rf{25062014-man-09x2}, \rf{25062014-man-09x3} , we see that the field $b$ can really be considered as the Nakanishi-Laudrup field.
Note also that, only for $d=4$, the field $b$ can be excluded from the Lagrangian by using equations of motion for $b$.

In the framework of AdS/CFT correspondence, the fields $\phi^a$, $\phi$ appear as boundary values of non-normalizable solution of equations of motion for spin-1 massless AdS field. This is the reason why we think that relation \rf{25062014-man-11}  can be considered as geometrical interpretation of the Nakanishi-Laudrup field.

\subsection{Spin-2 conformal field }

For spin-2 conformal field, Lagrangian \rf{25062014-man-07} takes the form
\beq
\label{25062014-man-23}  \LL_\tot  & = &  \frac{1}{4} \phi^{ab} \Box^{k+1}\phi^{ab} -
\frac{1}{8} \phi^{aa} \Box^{k+1}\phi^{bb}
+ \half \phi^a \Box^k \phi^a + \half \phi \Box^{k-1} \phi
\nonumber\\
& + &  \cb^a \Box^{k+1} c^a + \cb \Box^k c \,, \hspace{2cm} k \equiv \frac{d-2}{2}\,.
\eeq
Lagrangian \rf{25062014-man-23} is invariant under the BRST transformations
\beq
\label{25062014-man-24} && \ssf \phi^{ab} =     \partial^a c^b + \partial^b c^a + \frac{2}{d-2} \eta^{ab} c  \,,
\\
\label{25062014-man-25} && \ssf \phi^a =    \partial^a c - \Box c^a \,,
\\
\label{25062014-man-26} && \ssf \phi = -   u \Box c\,,
\\
\label{25062014-man-27} && \ssf c^a = 0 \,, \qquad \ssf c = 0 \,,
\\
\label{25062014-man-28} && \ssf \cb^a =     \partial^b \phi^{ab} - \half \partial^a \phi^{bb} + \phi^a  \,,
\\
\label{25062014-man-29} && \ssf \cb =    \partial^a \phi^a + \half \Box \phi^{aa} + u \phi \,, \qquad u \equiv \Bigl(2\frac{d-1}{d-2}\Bigr)^{1/2}\,.
\eeq
The anti-BRST transformations are obtained from  \rf{25062014-man-23}-\rf{25062014-man-29} by using the substitution $\ssf \rightarrow \ssfb$ and the following substitutions for all gauge fields and Faddeev-Popov fields: $\phi\rightarrow \phi$, $c\rightarrow \cb$, $\cb\rightarrow -c$.

It is easy to verify that BRST transformations \rf{25062014-man-23}-\rf{25062014-man-29}  are off-shell nilpotent. This is related to the fact that the vector and scalar fields $\phi^a$, $\phi$ can be realized as the Nakanishi-Laudrup field. To see this we introduce, in place of  the field $\phi^a$,  $\phi$,  new fields $b^a$,  $b$ by the following relations
\be \label{30062014-man-01}
b^a \equiv     \partial^b \phi^{ab} - \half \partial^a \phi^{bb} + \phi^a  \,, \qquad b  \equiv  \partial^a \phi^a + \half \Box \phi^{aa} + u \phi \,.
\ee
Using \rf{30062014-man-01} we get
\be \label{30062014-man-02}
\phi^a = b^a - \partial^b \phi^{ab} + \half \partial^a \phi^{bb}\,,\qquad u\phi = b - \partial^a b^a + \partial^a\partial^b \phi^{ab} - \Box \phi^{aa}\,.
\ee
Plugging $\phi^a$, $\phi$ \rf{30062014-man-02} into \rf{25062014-man-23}, we cast Lagrangian \rf{25062014-man-23} into the following form:
\beq
\label{30062014-man-03} && \LL_\tot = \LL + \LL_\qu\,,
\\
\label{30062014-man-04} && \LL = R_\lin^{ab} \Box^{k-1} R_\lin^{ab} - \frac{d}{4(d-1)} R_\lin\Box^{k-1}R_\lin \,,
\\
\label{30062014-man-05} && \LL_\qu = - b^a \Box^k (\partial^b \phi^{ab} - \half \partial^a \phi^{bb} ) + \frac{1}{u^2} (b -\partial^a b^a) \Box^{k-1}( \partial^c\partial^e \phi^{ce} - \Box \phi^{cc})
\nonumber\\
&& \hspace{1cm} + \half b^a \Box^k b^a + \frac{1}{2u^2} (b - \partial^a b^a) \Box^{k-1} (b - \partial^c b^c) + \cb^a \Box^{k+1} c^a + \cb \Box^k c\,, \qquad
\eeq
where $R_\lin^{ab}$, $R_\lin$ stand for the respective linearized Ricchi tensor and Ricchi scalar,
\beq
&& R_\lin^{ab} = \half ( -\Box\phi^{ab} + \partial^a \partial^c\phi^{cb} +
\partial^b \partial^c\phi^{ca} - \partial^a \partial^b \phi^{cc}),
\\
&& R_\lin = \partial^a\partial^b\phi^{ab}  - \Box \phi^{aa}\,.
\eeq
Using BRST transformations \rf{25062014-man-24}-\rf{25062014-man-29} and relations \rf{30062014-man-02}, we obtain
\be \label{30062014-man-06}
\ssf \cb^a =   b^a \,, \qquad \ssf \cb =  b\,, \qquad \ssf b^a = 0\,, \qquad \ssf b = 0\,.
\ee
From \rf{30062014-man-03},\rf{30062014-man-06}, we see that the fields $b^a$, $b$ can really be considered as Nakanishi-Laudrup fields.

\subsection{ Arbitrary spin-$s$ conformal field  }


{\bf BRST symmetries}. For spin-$s$ conformal field, a generating form of BRST invariant Lagrangian is given in \rf{25062014-man-07}. This Lagrangian is invariant under the following BRST and anti-BRST transformations:
\beq
\label{30062014-man-07} && \ssf |\phi\rangle  =     G \ck\,,\qquad \ssf \ck  =0\,, \hspace{1.7cm}  \ssf | \cb \rangle   =    \Lb  \phik  \,,
\\
\label{30062014-man-08} && \ssfb |\phi\rangle  =     G \cbk\,, \qquad \ssfb \ck  = -   \Lb  \phik\,, \qquad  \ssfb | \cb \rangle   =  0\,,
\eeq
where the operators $G$, $\Lb$ are given by
\beq
&& G \equiv    \alpar - e_1    -  \alpha^2\frac{1}{2N_\alpha +
d- 2} \eb_1\,,
\\
&& \Lb  \equiv \albpar - \half \alpar \bar\alpha^2  - \eb_1 \Pi^\smponetwo + \half
e_1 \bar\alpha^2  \,,
\\
&& e_1 =  \alpha^z e_z \Box\,, \qquad
\eb_1 =  - e_z \bar\alpha^z\,, \qquad  e_z = \Bigl(\frac{2s+d-4-N_z}{2s+d-4-2N_z}\Bigr)^{1/2}\,,
\eeq
while the operator $\Pi^\smponetwo$ is defined in Appendix. It is easy to verify that BRST and anti-BRST transformations \rf{30062014-man-07}, \rf{30062014-man-08}  are off-shell nilpotent. In the case under consideration, the Nakanishi-Laudrup field $\bk$ is defined by the relation $\bk\equiv \Lb\phik$.  Using this relation and \rf{30062014-man-07}, \rf{30062014-man-08}, we find the relations $\ssf \cb =\bk$, $\ssf \bk=0$, $\ssfb \ck = - \bk$, $\ssfb \bk=0$, which demonstrate that the ket-vector $\bk$ can really be considered as  the Nakanishi-Laudrup field.

To illustrate a structure of Lagrangian \rf{25062014-man-06}, we use the decomposition of the ket-vectors into scalar, vector, and tensor fields of the $so(d-1,1)$ algebra,
\beq
\label{28062014-man-06} && |\phi\rangle = \sum_{s'=0}^s \frac{\alpha_z^{s-s'}}{\sqrt{(s-s')!}}|\phi^{s'}\rangle\,, \qquad \ \ \ |\phi^{s'}\rangle\equiv \frac{1}{s'!}
\alpha^{a_1} \ldots \alpha^{a_{s'}} \phi^{a_1\ldots a_{s'}} |0\rangle\,.
\\
\label{28062014-man-07} && \ck = \sum_{s'=0}^{s-1} \frac{\alpha_z^{s-1-s'}}{\sqrt{(s-1-s')!}}|c^{s'}\rangle\,, \qquad |c^{s'}\rangle\equiv \frac{1}{s'!}
\alpha^{a_1} \ldots \alpha^{a_{s'}} c^{a_1\ldots a_{s'}} |0\rangle\,.
\\
\label{28062014-man-08} && \cbk = \sum_{s'=0}^{s-1} \frac{\alpha_z^{s-1-s'}}{\sqrt{(s-1-s')!}}|\cb^{s'}\rangle\,, \qquad |\cb^{s'}\rangle\equiv \frac{1}{s'!}
\alpha^{a_1} \ldots \alpha^{a_{s'}} \cb^{a_1\ldots a_{s'}} |0\rangle\,.
\eeq
Note that tensorial gauge fields in \rf{28062014-man-06} are double-traceless tensor fields of the $so(d-1,1)$ algebra, while tensorial Faddeev-Popov fields in  \rf{28062014-man-07}, \rf{28062014-man-08} are traceless tensor fields of the $so(d-1,1)$ algebra. Plugging ket-vectors \rf{28062014-man-06}-\rf{28062014-man-08} into \rf{25062014-man-06} we get the following representation for Lagrangian \rf{25062014-man-06}:
\beq
\LL_\tot & = & \sum_{s'=0}^s \LL^{s'} + \sum_{s'=0}^{s-1} \LL_{_\FP}^{s'}\,,
\nonumber\\
\LL^{s'} & = &  \frac{1}{2 s'!}\Bigl(
\phi^{a_1\ldots a_{s'}} \Box^{\nu_{s'}} \phi^{a_1\ldots a_{s'}}  - \frac{s'(s'-1)}{4} \phi^{aaa_3\ldots a_{s'}} \Box^{\nu_{s'}} \phi^{bba_3\ldots a_{s'}}\Bigr)\,,
\nonumber\\
\LL_{_\FP}^{s'} & = &  \frac{1}{s'!} \cb^{a_1\ldots a_{s'}} \Box^{\nu_{s'}+1} c^{a_1\ldots a_{s'}}\,,  \hspace{2cm}  \nu_{s'} = s'+\frac{d-4}{2}\,.
\eeq

\noindent {\bf Partition function}. For the computation of a partition function, it is convenient to decompose the double-traceless ket-vector $\phik$ into two traceless ket-vectors $|\phi_{_\I} \rangle$, $|\phi_{_\II} \rangle$,
\beq
\label{25062014-man-30} && \hspace{-1.5cm} \phik = |\phi_{_\I} \rangle   + \alpha^2 \NN |\phi_{_\II} \rangle\,, \qquad  \bar\alpha^2 |\phi_{_\I} \rangle =0\,,  \qquad \bar\alpha^2 |\phi_{_\II} \rangle =0\,,
\nonumber\\
&& \hspace{-1.5cm} \NN\equiv ((2s+d-4-2N_z)(2s+d-6-2N_z))^{-1/2}\,.
\eeq
Plugging \rf{25062014-man-30} into \rf{25062014-man-06}, we get
\be \label{25062014-man-31}
\LL_\tot =  \half \langle \phi_{_\I} | \Box^\nu  |\phi_{_\I} \rangle - \half \langle \phi_{_\II} | \Box^\nu  |\phi_{_\II} \rangle
+  \langle \cb |\Box^\nu  |c\rangle\,.
\ee
The decomposition of the ket-vectors $|\phi_{_{\I,\II}} \rangle$ into scalar, vector and tensor fields $\phi_{\I,\II}^{a_1\ldots a_{s'}}$  takes the form
\beq
\label{25062014-man-32} && |\phi_\I\rangle  = \sum_{s'=0}^{s} \frac{\alpha_z^{s-s'}}{\sqrt{(s-s')!}} |\phi_\I^{s'}\rangle \,, \qquad |\phi_\II\rangle  = \sum_{s'=0}^{s-2} \frac{\alpha_z^{s-2-s'}}{\sqrt{(s-2-s')!}} |\phi_\II^{s'}\rangle \,,
\nonumber\\
&& |\phi_{\I,\II}^{s'}\rangle = \frac{1}{s'!}  \alpha^{a_1} \ldots \alpha^{a_{s'}}  \phi_{\I,\II}^{a_1\ldots a_{s'}}|0\rangle\,.
\eeq
Note that the fields $\phi_{\I,\II}^{a_1\ldots a_{s'}}$ are totally symmetric traceless tensor fields of the $so(d-1,1)$ algebra.
In terms of the scalar, vector, and traceless tensor fields, Lagrangian \rf{25062014-man-31} can be represented as
\beq
\label{25062014-man-33} &&   \LL_\tot  = \sum_{s'=0}^s \LL_\I^{s'} - \sum_{s'=0}^{s-2} \LL_\II^{s'} +
\sum_{s'=0}^{s-1} \LL_\FP^{s'}\,, \qquad
\\
\label{25062014-man-34} &&     \LL_\I^{s'}  = \frac{1}{2s'!}\, \phi_\I^{a_1 \ldots a_{s'}} \Box^{\nu_{s'}}  \phi_\I^{a_1 \ldots a_{s'}}\,,
\qquad   \LL_\II^{s'}  =  \frac{1}{2s'!}\, \phi_\II^{a_1 \ldots a_{s'}} \Box^{\nu_{s'}+2}    \phi_\II^{a_1 \ldots a_{s'}}\,,
\\
\label{25062014-man-36} &&   \LL_\FP^{s'}  = \frac{1}{s'!}\, \cb^{a_1 \ldots a_{s'}} \Box^{\nu_{s'}+1}   c^{a_1 \ldots a_{s'}}\,.
\eeq
From \rf{25062014-man-33}-\rf{25062014-man-36}, we see that the partition function is given by
\beq
\label{25062014-man-37} && Z = \prod_{s'=0}^{s-1} (D^{s'} D^{s'})^{\nu_{s'}+1} \Bigr/ \prod_{s'=0}^s (D^{s'})^{\nu_{s'}}  \prod_{s'=0}^{s-2} (D^{s'})^{\nu_{s'}+2}\,,
\\
\label{25062014-man-38} && D^{s'} \equiv \bigl(\det (-\Box ) \bigr)^{1/2}\,,
\eeq
where, in relation \rf{25062014-man-38}, determinant of  the Laplace operator is evaluated on space of rank-$s'$ traceless tensor field.
It is easy to see that partition function \rf{25062014-man-37} can be simplified as follows
\be \label{26062014-man-01}
Z = \frac{(D^{s-1})^{\nu_s + 1}}{(D^s)^{\nu_s}}\,, \qquad \nu_s = s + \frac{d-4}{2}\,.
\ee

Partition function \rf{26062014-man-01} was obtained by different method in earlier literature in Refs.\cite{Fradkin:1985am,Tseytlin:2013jya}). Also, in Ref.\cite{Tseytlin:2013jya} it was noted that it is helpful to use the following relation %
\be \label{26062014-man-09z1}
D^{s'} = D^{s'\perp} D^{s'-1},
\ee
where $D^{s'\perp}$ takes the same form as in \rf{25062014-man-38}, while the determinant  of Laplace operator is evaluated on space of traceless and divergence-free rank-$s'$ tensor field. Namely, by using \rf{26062014-man-09z1}, partition function \rf{26062014-man-01} can be cast into the form
\be \label{30062014-man-14}
Z = \frac{1}{(D^{s\perp})^{(d-4)/2}} \prod_{s'=0}^{s-1} \frac{D^{s'\perp}}{D^{s\perp}}\,.
\ee
In Ref.\cite{Tseytlin:2013jya}, expression \rf{30062014-man-14} was generalized to a partition function for a conformal field in (A)dS space. Computation of the partition function for the conformal field in (A)dS by various methods may be found in Refs.\cite{Tseytlin:2013jya,Metsaev:2014iwa}.

\bigskip
{\bf Acknowledgments}. This work was supported by the RFBR Grant No.14-02-01171.

\setcounter{section}{0}\setcounter{subsection}{0}
\appendix{ \large Notation and conventions  }

The vector indices of the $so(d,1)$ algebra take the values $A,B,C,E=0,1,\ldots ,d$, while the vector indices of the $so(d-1,1)$ algebra take the values   $a,b,c,e=0,1,\ldots ,d-1$. To simplify our expressions we drop the flat metrics  $\eta^{AB}=(-,+,\ldots,+)$ and $\eta^{ab}=(-,+,\ldots,+)$ in scalar product, i.e., we use $X^AY^A \equiv \eta_{AB}X^A Y^B$,
$X^aY^a \equiv \eta_{ab}X^a Y^b$.

Covariant derivative with flat indices $D^A$ is defined by the relations $D^A = \eta^{AB}D_B$,
\be  \label{29062014-man-01}
D_A \equiv e_A^\mun D_\mun\,,  \hspace{0.5cm} D_\mun \equiv
\partial_\mun + \half \omega_\mun^{AB} M^{AB}\,, \hspace{0.5cm} M^{AB} = \alpha^A \bar\alpha^B - \alpha^B \bar\alpha^A\,,
\ee
$\partial_\mun = \partial/\partial x^\mun$, $\mun = 0,1,\ldots, d$, where $x^\mun$ are base manifold coordinates of (A)dS space, $e_A^\mun$ is inverse of vielbein  $e_{\mun A}$, $e_A^\mun e_{\mun B} = \eta_{AB}$,
$D_\mun$ covariant derivative with Lorentz connection $\omega_\mun^{AB}$, and  $M^{AB}$ is spin operator of Lorentz algebra $so(d,1)$.
Contravariant tensor field in  (A)dS with flat indices, $\Phi^{A_1\ldots A_s}$, is related to contravariant tensor field with base manifold indices, $\Phi^{\mun_1\ldots \mun_s}$, in a standard way: $\Phi^{A_1\ldots A_s} \equiv e_{\mun_1}^{A_1}\ldots e_{\mun_s}^{A_s} \Phi^{\mun_1\ldots \mun_s}$. D'Alembert operator in (A)dS is defined by the relation
\be
\Box_{_\pAdS}   \equiv D^AD^A + \omega^{AAB}D^B\,, \qquad \omega^{ABC} \equiv e^{A \mun}\omega_\mun^{BC}\,,\qquad e \equiv \det e_\mun^A\,.
\ee

Derivative $D_\mun$ \rf{29062014-man-01} is defined in space of ket-vectors constricted out of $\alpha^A$. Derivative acting on tensor field are denoted by  $\DD_\mun$.  Actions of such derivative on vector field with flat indices is defined in a standard way
\be
\DD_\mun \phi^A = \partial_\mun \phi^A + \omega_\mun^{AB}(e)\phi^B\,.
\ee
In place of $\DD_\mun$, we prefer to use a derivative with flat indices, $\DD^A$,
\be
 \DD_A \equiv e_A^\mun \DD_\mun\,,\qquad \DD^A = \eta^{AB}\DD_B\,,
\qquad [\DD^A,\DD^B] \phi^C = R^{ABCE}\phi^E \,,
\ee
$\DD^2\equiv \DD^A \DD^A$, where Riemann tensor of (A)dS space is given by
\beq
&& R^{ABCE} = \rho (\eta^{AC}\eta^{BE} - \eta^{AE}\eta^{BC})\,, \qquad \rho = \frac{\epsilon}{R^2}\,, \qquad \epsilon = \left\{ \begin{array}{cl} 1 & \hbox{for \ dS}
\\[5pt]
-1 & \hbox{for \ AdS}
\end{array}\right.\qquad
\eeq

Creation operators $\alpha^A$, $\zeta$ and the respective annihilation operators $\bar\alpha^A$, $\bar\zeta$  are referred to as oscillators in this paper. Commutaion relations, the vacuum $|0\rangle $, and hermitian conjugation rules are fixed by the relations
\be
[ \bar\alpha^A,\alpha^B] = \eta^{AB}, \quad [\bar\zeta,\zeta]=1,  \quad \bar\alpha^A |0\rangle = 0\,,\quad \bar\zeta |0\rangle = 0\,,   \quad
\alpha^{A \dagger} = \bar\alpha^A\,, \quad \zeta^\dagger = \bar\zeta\,.
\ee
Oscillators $\alpha^A$, $\bar\alpha^A$ and  $\zeta$, $\bar\zeta$, transform in the respective vector and scalar representation of the $so(d,1)$ algebra. Derivatives with respect to space-time coordinates $x^a$, $z$ are denoted by $\partial^a \equiv \eta^{ab}\partial/\partial x^b$, $\partial_z \equiv \partial/\partial z$. In basis of the  $so(d-1,1)$ algebra, we use the decompositions $\alpha^A = \alpha^a, \alpha^z$ and $\eta^{AB}=\eta^{ab},\eta^{zz}$, where $\eta^{zz}=1$.  We adopt the following notation for the scalar product of oscillators and derivatives
\beq
\label{30062014-man-09} && \alphabf \Dbf \equiv \alpha^A D^A\,,\qquad \bar\alphabf \Dbf
\equiv \bar\alpha^A D^A\,,  \qquad \alphabf^2 \equiv \alpha^A \alpha^A\,,\qquad  \bar\alphabf^2 \equiv
\bar\alpha^A \bar\alpha^A\,,\qquad
\\
\label{30062014-man-10} && \alpha \partial \equiv \alpha^a\partial^a\,,\qquad \ \ \  \bar\alpha \partial
\equiv \bar\alpha^a\partial^a\,, \qquad \ \ \ \ \ \alpha^2 \equiv \alpha^a \alpha^a\,,\qquad \ \ \bar\alpha^2 \equiv
\bar\alpha^a \bar\alpha^a\,,
\\
\label{30062014-man-11} &&  N_\alphabf \equiv \alpha^A \bar\alpha^A  \,, \qquad \ \ N_\alpha \equiv \alpha^a \bar\alpha^a  \,, \qquad \ \ \ N_\zeta \equiv \zeta \bar\zeta \,, \qquad  \ \ \ \ N_z\equiv \alpha^z \bar\alpha^z \,,
\\
\label{30062014-man-12} &&  \Pi^\smponetwo \equiv 1 - \alpha^2\frac{1}{2(2N_\alpha + d)}\bar\alpha^2\,,\qquad  \Pibf^\smponetwo \equiv 1 - \alphabf^2\frac{1}{2(2N_\alphabf + d+1)}\bar\alphabf^2\,,
\\
&&   \Box \equiv \partial^a \partial^a \,, \qquad \mu \equiv 1- \frac{1}{4}\alpha^2 \bar\alpha^2  \,, \qquad \mubf \equiv 1-\frac{1}{4}\alphabf^2 \bar\alphabf^2\,.
\eeq
Bra-vectors and ket-vectors of Faddeev-Popov fields satisfy the relations $\cbr = \ck^\dagger$, $\cbbr = -\cbk^\dagger$, while Faddeev-Popov scalar, vector, and tensor fields satisfy the following hermitian conjugation conditions: $c^{A_1\ldots A_{s'}\dagger} = c^{A_1\ldots A_{s'}}$, $\cb^{A_1\ldots A_{s'}\dagger} = - \cb^{A_1\ldots A_{s'}}$ . To illustrate these hermitian conjugation rules we consider  Faddeev-Popov vector fields and note that, if ket-vectors are given by $\ck=\alpha^A c^A|0\rangle$, $\cbk=\alpha^A \cb^A|0\rangle$, then the respective bra-vectors
are given by $\cbr =\langle 0| \bar\alpha^A c^A$, $\cbbr = \langle 0|\bar\alpha^A \cb^A$.

\end{document}